\documentclass{aa}
\usepackage{graphicx}

\newcommand{\prp}    {${\rlap.}^{\prime}$}
\newcommand{\grp}    {${\rlap.}^{\circ}$}
\newcommand{\pri}    {${\rlap.}^{\prime \prime}$}
\newcommand{\rl}     {${\rlap.}^{s}$}

\newcommand{\ltsima} {$\; \buildrel < \over \sim \;$}
\newcommand{\simlt}  {\lower.5ex\hbox{\ltsima}}            
\newcommand{\gtsima} {$\; \buildrel > \over \sim \;$}
\newcommand{\simgt}  {\lower.5ex\hbox{\gtsima}}            

\begin{document}

\title{Possible hot spots excited by the relativistic jets of Cygnus X-3}

\author{
J. Mart\'{\i}\inst{1}
\and D. P\'erez-Ram\'{\i}rez\inst{1}
\and J.~L. Garrido\inst{1}
\and P. Luque-Escamilla\inst{2}
\and J.~M. Paredes\inst{3}
}

\offprints{J. Mart\'{\i}}

\institute{Departamento de F\'{\i}sica, EPS,  
Universidad de Ja\'en, Campus Las Lagunillas s/n, Edif. A3, 23071 Ja\'en, Spain \\
\email{jmarti@ujaen.es, dperez@ujaen.es, jlg@ujaen.es} 
\and Dpto. de Ing. Mec\'anica y Minera, EPS,
Universidad de Ja\'en, Campus Las Lagunillas s/n, Edif. A3, 23071 Ja\'en, Spain \\
\email{peter@ujaen.es}
\and Departament d'Astronomia i Meteorologia, 
Universitat de Barcelona, Av. Diagonal 647, 08028 Barcelona, Spain \\
\email{jmparedes@ub.edu}
}

\date{Received / Accepted}

\titlerunning{Possible hot spots excited by Cygnus X-3}

\abstract{We present the results of a deep search for associated radio
features in the vicinity of the microquasar Cygnus X-3. The motivation
behind is to find out evidence for interaction between its
relativistic jets and the surrounding interstellar medium, which could
eventually allow us to perform calorimetry of the total energy
released by this microquasar during its flaring lifetime. Remarkably,
two radio sources with mJy emission level at centimeter wavelengths
have been detected in excellent alignment with the position angle of
the inner radio jets. We propose that these objects could be the hot
spots where the relativitic outflow collides with the ambient gas in
analogy with Fanaroff-Riley II radio galaxies.  These candidate hot
spots are within a few arc-minutes of Cygnus X-3 and, if physically
related, the full linear extent of the jet would reach tens of
parsecs. We discuss here the evidence currently available to support
this hypothesis based on both archival data and our own observations.

}

\maketitle

\section{Introduction} 

Cygnus X-3 is a well known microquasar in the Galaxy, i.e., a X-ray
binary producing bipolar relativistic jets.  The neutron star or black
hole nature of the compact object in the binary system is still a
matter of discussion, but a consensus has been reached about the
normal companion being a WN Wolf-Rayet star (Fender et
al. \cite{f99}). The source is seen through an absorption of at least
$A_V = 10$ magnitudes and possibly higher.  The relativistic outflow
of this microquasar has been observed at both arc-second and
sub-arcsecond scales, with an ejection axis very close to the
North-South direction (see e.g. Mart\'{\i} et al. \cite{marti01} and
Miller-Jones et al. \cite{miller04}).  The reader is also referred to
these references for a basic description of the main properties of
this system.  Here, we focus our attention on the unsolved problem of
finding evidence for the expected hot spot/double lobe structures at
the jet terminal shocks.  The absence of observational evidence for
such terminal features is an open question and not only for Cygnus X-3
itself but in a general context as well.  Galactic sources of
relativistic jets or microquasars pump significant amounts of energy
into the interstellar medium (ISM).  How such relativistic flows
affect the galactic ISM remains, unfortunately, difficult to observe
and study.  As an illustrative example, simple power estimates by
Fender \& Pooley (\cite{fender00}) suggest that $\sim 10^{51}$ erg
(i.e. comparable with a supernova explosion) are injected by the
superluminal microquasar \object{GRS 1915+105} into its surroundings
over its total ejecting life time.

Only in few cases we do see evidence of interaction between the jets
and its surroundings, as in the double lobes of the Galactic Center
microquasars \object{1E~1740.7$-$2942} (Mirabel et al. \cite{mir92})
and \object{GRS 1758$-$258} (Rodr\'{\i}guez et al. \cite{lf92}), or a
clear deceleration against the ISM as in the remarkable case of
\object{XTE~J1550$-$564} (Corbel et al. \cite{cor02}).  Recently, a
possible extended bow shock with arc-shaped morphology has been also
reported at radio wavelengths for the microquasar \object{Cygnus X-1}
(Gallo et al. \cite{gal05}).

The known population of galactic microquasars has grown significantly
in recent times up to a total of at least 15 systems currently
catalogued (Paredes \& Mart\'{\i} \cite{pm03}).  Nevertheless, the
observational evidence of interaction between their relativistic jets
and the ambient gas continues being the exception rather than the
rule.  Indeed, one would expect to see several examples of hot
spot/double lobe structures at the jet terminal shocks in a process
similar to extragalactic Fanaroff-Riley type II (FRII) sources
(Fanaroff \& Riley \cite{fr74}). In contrast, most relativistic jets
in the Galaxy fade out before they significantly decelerate when their
ram pressure equals that of the ISM.

This situation is indeed puzzling specially when comparing these
relativistic jet sources with the thermal jets from young stellar
objects (YSOs). Jets from YSOs are orders of magnitude less powerful
than those of microquasars, but yet they are able to create a wide
variety of bow shock structures.  These are the well known Herbig-Haro
objects whose latest catalog contains a few hundreds of entries
(Reipurth \cite{bo99}). Why such differences in number exist?.  Are
YSOs bow shocks easier to observe only because they decelerate in a
denser medium such as a molecular cloud with particle densities $\sim
10^{4}$ cm$^{-3}$?.  Or is there an additional observational
bias?. Heinz (\cite{he02}) has argued that microquasars are typically
located in lower density environments, in a dynamical sense, than
their AGN relatives. Tenuous media with densities well below the
canonical $\sim 1$ cm$^{-3}$ would certainly imply the absence of
interaction signatures, but additional work is needed to test whether
this holds for most systems.

Early searches for large-scale structures connected with GRS 1915+105
revealed the existence of two intriguing IRAS sources whose radio
emission is mostly of thermal free-free nature. They are symmetrically
located with respect to GRS 1915+105 and perfectly aligned with the
position angle of its relativistic jets (Rodr\'{\i}guez \& Mirabel
\cite{lf98}; Chaty et al. \cite{ch01}).  A recent revision of their
observational properties by Kaiser et al. (\cite{kaiser04}) concludes
that it is perfectly conceivable, both energetically and dynamically,
that the two IRAS sources are the jet impact regions on the ISM at a
distance of tens of parsecs from their exciting source.  Shock waves
and turbulent plasma motion in the terminal lobes would naturally lead
to synchrotron non-thermal emission explaining the unusual non-thermal
feature in a otherwise normal H{\small II} region.  Their conclusions
are based on applying a scaled down model for large-scale structure of
jets in extragalactic radio sources (Kaiser \& Alexander
\cite{kaiser97}).  The average energy transport rate needed to sustain
the proposed double structure is relatively low ($\sim 10^{36}$
erg~s$^{-1}$), with an estimated age of $\sim 1$ Myr.  If the IRAS
sources association with GRS 1915+105 is confirmed, that would place
this microquasar at a much lower distance than usually assumed.

In the case of the microquasar Cygnus X-3, a similar pair of symmetric
radio sources aligned with the jets position angle was proposed by
Mart\'{\i} et al. (\cite{marti00}). However, none of them displayed
possible non-thermal filaments pointing towards the central core of
Cygnus X-3 as in the GRS 1915+105 case. Moreover, their separation was
exceedingly large ($\sim 100$ pc) for a physical connection to be
claimed without further evidence.  All these facts induced us to
carefully look for fainter and closer radio sources, not previously
noticed, which could be associated with the missing large scale radio
lobes.  In this paper, we report the results of this search showing
that Cygnus X-3 is likely another microquasar system where large-scale
deceleration/interaction effects of the jets can be observed.

\section{Discovery of Cygnus X-3 hot spot candidates}

\begin{table*}
\caption[]{\label{obsrad} Radio properties of possible hot spots associated with Cygnus X-3 derived from VLA observations}
\begin{tabular}{ccccccc}
\hline
\hline
Hot spot    & Position                                           & Frequency  &  Peak Flux Density$^*$ & Integrated Flux Density$^*$ & Spectral \\
candidate   & $\alpha_{{\rm J2000.0}}$, $\delta_{{\rm J2000.0}}$ &   (GHz)    &         (mJy)          &           (mJy)             & Index    \\
\hline
            &                                                 &       &               &               &                            \\
North       & $20^h 32^m$26\rl 88$\pm$0\rl 01                 & 1.425 & $1.63\pm0.13$ & $3.3\pm0.5$   &  $-0.4\pm0.2$              \\
            & $+41^{\circ} 04^{\prime}$32\pri 9$\pm$0\pri 2   & 4.860 & $1.25\pm0.11$ & $1.9\pm0.4$   &                            \\
            &                                                 &       &               &               &                            \\
South       & $20^h 32^m$24\rl 97$\pm$0\rl 02                 & 1.425 &    $\leq 0.6$ &     $-$       &  $\geq -0.7$$^{**}$ \\
            & $+40^{\circ} 53^{\prime}$05\pri 9$\pm$0\pri 2   & 4.860 & $0.27\pm0.04$ & $0.45\pm0.09$ &                            \\
            &                                                 &       &               &               &                            \\
\hline
\hline
\end{tabular}
~\\
$^*$ Corrected for primary beam decay using the AIPS task PBCOR but not for bandwidth smearing. \\
$^{**}$ Computed using the peak and $4\sigma$ upper limit values corrected for bandwidth smearing. \\
\end{table*}

With the idea of searching for faint radio emission around Cygnus X-3
we have reanalyzed different observations retrieved from the archive
of the Very Large Array (VLA) of NRAO in New Mexico (USA).  The
results reported here come mainly from three different VLA projects,
namely those with identification codes AM551, AS483 and AM669. All of
them were reduced using the AIPS package of NRAO applying standard
procedures for calibration of interferometer data. All projects used
3C286 as the amplitude calibrator and 2007+404 as the phase
calibrator, at an angular distance of 4\grp 7 from Cygnus X-3.

The first VLA project AM551 was conducted in 1997 by Mart\'{\i} et
al. (\cite{marti00}) with the array in B configuration at the 6 cm
wavelength being, in principle, sensitive to angular scales up to
$\sim 30^{\prime\prime}$.  Its main result was the detection of
transient arc-second radio jets of Cygnus X-3. Our hope when
reanalyzing AM551 was that its deep integration time ($\sim 8$~h)
could eventually reveal additional features fainter and/or further
away from the microquasar core than those originally discovered.  At
the time of observations, Cygnus X-3 was in a ``quiescent'' state
($S_{{\rm 6cm}} \sim 100$ mJy) and its variable radio core was
subtracted with a similar approach as described by Mart\'{\i} et
al. (\cite{marti00}) in order to remove artifacts induced by the
microquasar variability during the CLEAN deconvolution process.

\begin{figure*}
   \centering
\resizebox{\hsize}{!}{\includegraphics{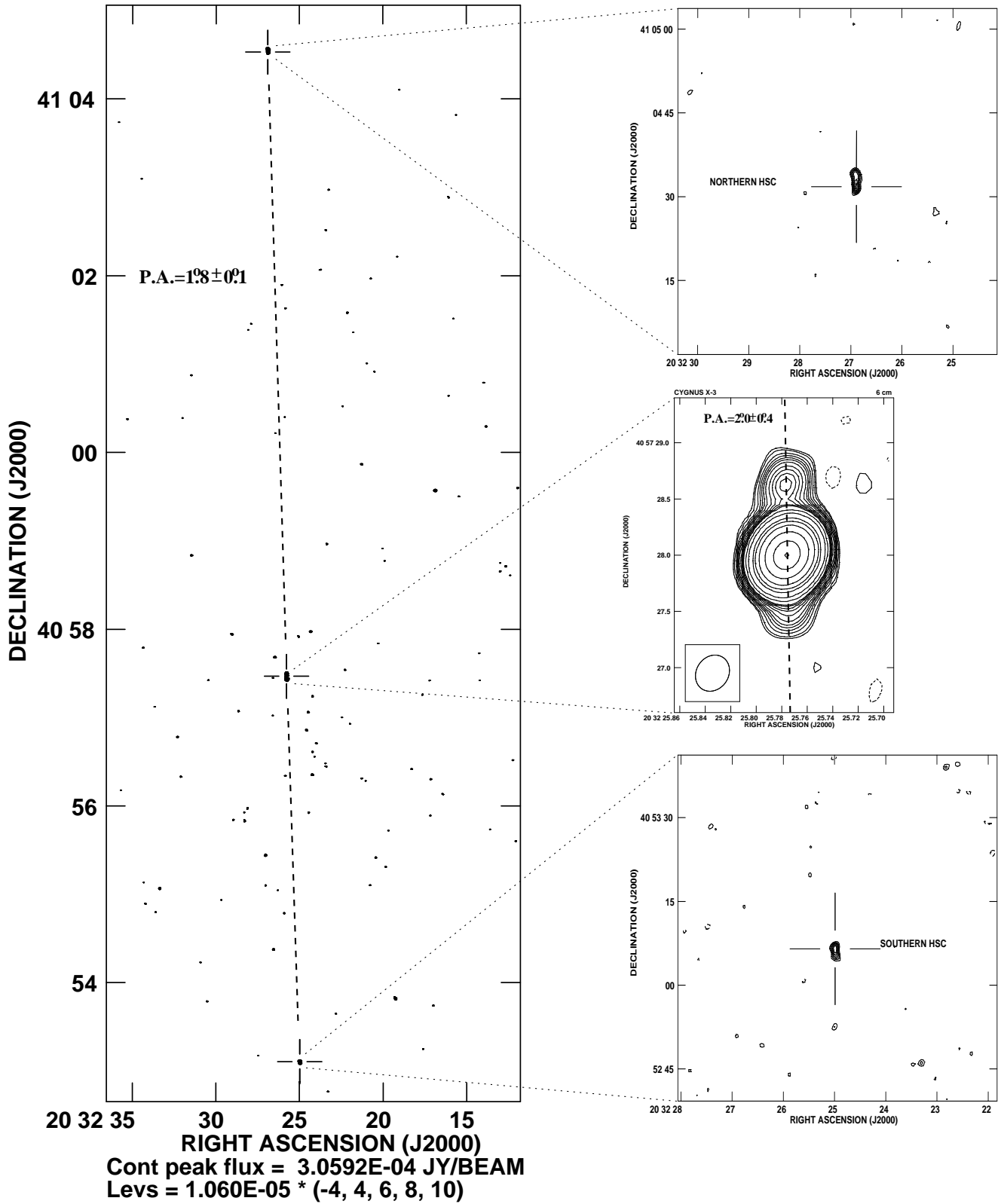}}
      \caption{{\bf Left.} Natural weight map of the
Cygnus X-3 field from a re-analysis of VLA project AM551.
The original observation was conducted on 4 May 1997
at the 6~cm wavelength in B configuration.
The map covers almost the totality of the antenna primary beam and remarkably shows
the existence of two radio sources perfectly aligned with the position angle of the
Cygnus X-3 inner radio jets. These are the only objects in the field with peak flux density above
four times the rms noise of 10.6 $\mu$Jy beam$^{-1}$, in addition to Cygnus X-3 itself. 
We interpret them as the possible hot spots
where a larger scale flow terminates. The clean components have been restored 
using an almost circular Gaussian beam of 1\pri 5. 
{\bf Right.} The top and bottom panels are zoomed contour plots of the two hot spot candidates. 
Their morphology is difficult to judge because of severe bandwidth and time smearing
at their distance from the phase center, although they could be slightly resolved. 
The central panel shows the Cygnus X-3 arc-second radio jets taken from
Mart\'{\i} et al. \cite{marti01} for illustrative purposes and corresponding to a strong
radio outburst in late 2000.}
      \label{vla_panel}
   \end{figure*}

The resulting image is shown in the left panel of
Fig. \ref{vla_panel}.  Here, two faint radio sources are clearly
detected at angular distances of 7\prp 07 and 4\prp 36 from Cygnus X-3
and roughly in the North-South direction. Interestingly, they are in
almost perfect alignement with Cygnus X-3 along a position angle of
1\grp 8$\pm$0\grp 1.  This value agrees well with the position angle
of the Cygnus X-3 inner arc-second radio jets, measured to be 2\grp
0$\pm$0\grp 4 by Mart\'{\i} et al. (\cite{marti01}) Moreover, the
asymmetry between the separation of the northern and southern sources
is reminiscent of the asymmetry also existing between the northern and
southern inner jet components (see the central right panel of
Fig. \ref{vla_panel}). Such asymmetry could be accounted for if the
northern and southern jets decelerate in regions of moderately
different ISM density.

The observational properties of these newly discovered radio sources
are summarized in Table \ref{obsrad}. Wide field and zoomed contour
maps are also presented in the different panels of
Fig. \ref{vla_panel}. At their distances from the phase center, the
new sources suffer from significant bandwidth smearing giving them an
elongated shape poiting towards Cygnus X-3, specially in the northern
case.  Such instrumental limitation makes very difficult to assess
which is their intrinsic morphology and this is the reason why angular
diameters are not included in Table \ref{obsrad}. Nevertheless, the
possibility that the sources are slightly extended cannot be ruled out
based on present data.  Indeed the slightly curved shape of the
northern object is not strictly consistent with a pure radial smearing
of a compact source.  A crude estimate of its transversal angular size
can be obtained by taking the deconvolved minor axis of an elliptical
Gaussian fit using the AIPS task IMFIT.  This size is about 1\pri 6
and we will later adopt an aspect ratio of 1 for discussion purposes.

From source counts (see Rodr\'\i guez et al. \cite{lf89}) and simple
geometric arguments, we estimate that the probability of randomly find
two background radio sources brighter than 1 mJy, aligned within $\sim
1^{\circ}$ of the jet position angle and closer than 7\prp 0 from
Cygnus X-3, is of order $\sim 10^{-5}$ percent. Indeed, no other pair
of radio sources but these two appear so well aligned in the field.

In order to estimate the source's spectral index $\alpha$ (defined as
$S_{\nu} \propto \nu^{\alpha}$, where $S_{\nu}$ is the radio flux
density at a given frequency $\nu$), and hence constraining the
emission mechanism, we searched the VLA archives for suitable
observations at a different wavelength.  The VLA project AS483
conducted in year 1993 by A. Smale in B configuration was found to be
appropriated for such purpose. A re-analysis of it provided a clear
detection of the northern radio source, at the 20~cm wavelength, and
an upper limit to the brightness of the southern one.  The 20~cm
results and the resulting spectral index information are also included
in Table \ref{obsrad}.  Using the Smale data, the radio spectrum of
the northern candidate is well described by $S_{\nu} = 3.9\pm0.4~{\rm
mJy}~[\nu /{\rm GHz}]^{-0.4\pm0.2}$, which clearly suggests a
non-thermal synchrotron emission mechanism. The spectral index of the
southern source is also consistent with a similar negative value, i.e,
typical of optically thin synchrotron radiation.

Taking together i) the almost perfect alignment with the jet's
position angle, ii) the north/south asymmetry similar to that of the
arc-second radio jets, and iii) the likely non-thermal emission of the
two radio sources discovered, we propose them as serious candidates to
be the impact sites of the Cygnus X-3 jets against the ISM. Hereafter,
we will refer to them as the northern and southern hot spot candidates
(HSCs).

\section{Setting upper limits on proper motion}

Following their discovery, the two HSCs were also detected when
carefully inspecting very wide field VLA maps obtained from the AM669
data.  This project was carried out during late 2000 with the array
being in its most extended A configuration and published by Mart\'{\i}
et al. (\cite{marti01}).  The new AM669 maps (not shown here) also
suffer from noticeable smearing like those of
Fig. \ref{vla_panel}. However, both the northern and southern hot spot
candidates were detected although with low signal-to-noise ratios of 5
and 4, respectively.  Using the longer time baseline available, 7.6 yr
between the AM669 and AS483 data, we can try to set an upper limit to
the proper motion of the north HSC with respect to Cygnus X-3.  The
total projected length of the northern flow $L_j \sin{\theta}/D$ would
be equivalent to 7\prp 07, where $\theta$ is the angle with the line
of sight and $D$ the distance to the source.

Between 1993 and 2000, the north HSC shifted its separation from
Cygnus X-3 by less than about 0\pri 7. This value is comparable or
smaller than one synthesized beam for any of the two observing epochs
involved.  Considering that smearing is also likely to affect our 6~cm
astrometry at large distances from the phase center, we conservatively
consider that no reliable proper motion can be claimed based on the
present data.  Therefore, a safe upper limit on the advance speed of
the end of the possible jet can be expressed as $\dot{L}_j \leq
4400~{\rm km~s}^{-1} D_{10} (\sin{\theta})^{-1}$ being $D_{10}$ the
source distance in units of 10 kpc.

\section{Search for X-ray, optical and infrared counterparts}

In order to exclude the possible stellar or extragalactic nature for
the HSCs, their accurate radio positions in Table \ref{obsrad} have
been searched at X-ray, optical and infrared wavelengths.

\subsection{X-rays}

We have inspected the HSCs location in several X-ray mission archives,
including ROSAT, XMM and CHANDRA. We only found a ROSAT PSPC image
(0.76 -- 2.04 KeV band) from which an upper limit count rate of 6.245
counts s$^{-1}$ was inferred.  Assuming a photon index $\Gamma = 1.5$
and an absorption of $A_V = 10$ mag towards Cygnus X-3, equivalent to
$N_H = 1.8 \times 10^{22}$ cm$^{-2}$ (Predehl \& Schmitt
\cite{predehl95}), the corresponding unabsorbed X-ray flux limit is
225 $\mu$Jy.

\subsection{Optical observations}

In the optical domain, we observed the two positions with the 2.2~m
telescope at the Centro Astron\'omico Hispano Alem\'an (CAHA) in Calar
Alto (Spain).  We obtained deep $I$-band images using the SITe-d15 CCD
detector of the CAFOS instrument.  Astrometry of the CCD frames was
determined through twenty stars in the field whose positions were
retrieved using the {\it SkyCat} tool of the European Southern
Observatory.  The plate solutions were established by means of the
AIPS task XTRAN with a third order polinomial fit, whose residuals
were typically of 0\pri 5.

\begin{figure*}
   \centering
\resizebox{\hsize}{!}{\includegraphics{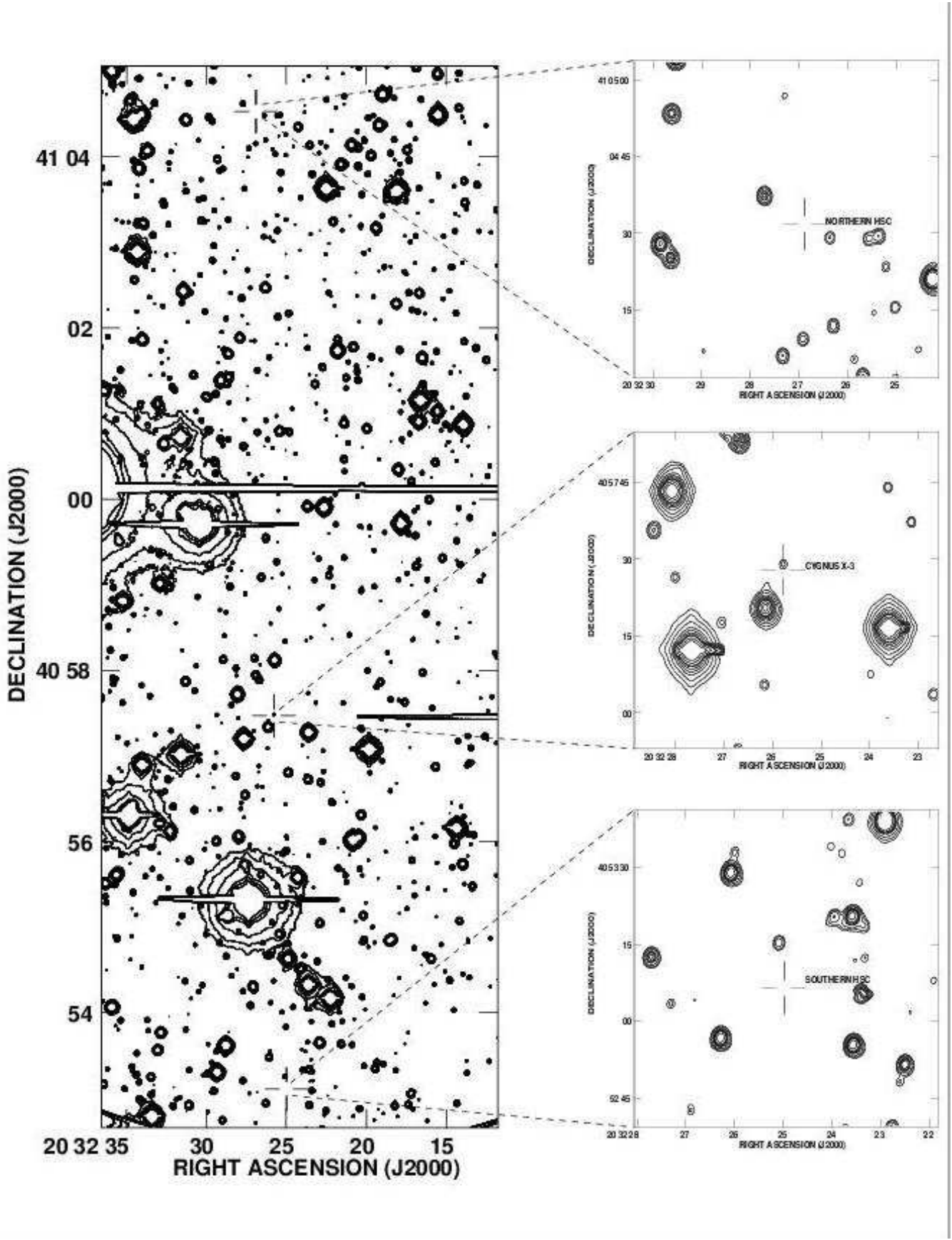}}
      \caption{CAHA 2.2~m $I$-band image taken on 2004 July 20 with the CAFOS instrument.
The crosses with a gap mark both the position of Cygnus X-3
and the two hot spot candidates. The right panels display an enlarged view of their
respective locations.}
      \label{caha_panel}
   \end{figure*}

Fig. \ref{caha_panel} shows the results of our CAHA observations and
astrometry.  Cygnus X-3 was clearly detected in the $I$-band with an
estimated magnitude of about $I=21.3$, which is comparable to previous
detections reported in the literature (Wagner et al. \cite{w89}). In
contrast, no optical counterpart was detected at the likely hot spot
location up to a limit magnitude of $I=22.4$.

\subsection{Infrared search}

The search for a counterpart at near-infrared wavelengths is strongly
justified given the strong interstellar absorption in the Cygnus X-3
region. The 2 Micron All Sky Survey (2MASS) was also inspected at the
position of the HSCs.

A near-infrared counterpart was not found for none of them within
$4^{\prime \prime}$ of their radio position. The corresponding upper
limit magnitudes are 17.8, 16.8 and 16.3 for the $J$, $H$ and $Ks$
bands, respectively.

\section{Discussion}

\begin{figure}
   \centering
\resizebox{\hsize}{!}{\includegraphics[angle=-90]{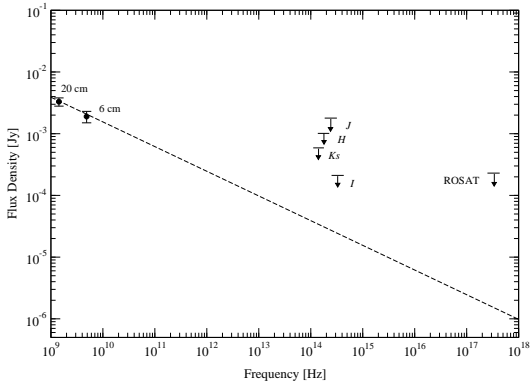}}
      \caption{Spectral energy distribution for the northern HSC expected
at infrared and X-ray wavelengths extrapolating a synchrotron spectrum with the same power law
index as derived from radio observations (dashed line).
The filled triangles represent upper limits for the optical (CAHA), near-infrared (2MASS)
and X-ray (ROSAT) emission after reddening correction assuming $A_V=10$ mag.
Higher absorption values, certainly not excluded for this source, would just rise
the upper limits plotted here not affecting the conclusions of our paper. 
}
      \label{sed}
   \end{figure}

The observed radio spectrum and infrared and X-ray upper limits are
consistent with extrapolation of the same spectral index derived from
20 and 6~cm observations (see Fig. \ref{sed}).  In the previous
sections, we have failed to identify an optical and near-infrared
counterpart with stellar appearance for the HSCs.  Had they been
background or foreground stars, or nearby AGNs, we would expect to
detect some optical or infrared source with compact appearance unless
absorption prevents us from doing so. In any case, the possibility
that we are not dealing with stars or AGNs remains clearly open.

Thus, we will proceed hereafter under the assumption that both HSCs
are physically related to Cygnus X-3 although this point needs yet to
be confirmed by further more sensitive observations.

During our discussion we will apply some of the equations used by
Kaiser et al. (\cite{kaiser04}) when revising the possible large scale
structure of GRS~1915+105.  In their work, thermal bremsstrahlung
radio emission is detected from the shock compressed ISM (i.e. the
extended IRAS sources).  Unfortunately, our application of their
formalism will necessarily be limited because we did not find evidence
of such thermal emission downstream of the HSC. The $4\sigma$ upper
limits at 6 cm are 0.6 and 0.1 mJy beam$^{-1}$ for the northern and
southern one, respectively.  The possibility that such emission exists
is, however, not ruled out by our present VLA observations due to lack
of sensitivity so far away from the phase center and the blindness of
the VLA B configuration array to angular scales larger than half
arc-minute.

It is instructive to assess here the possibility for Cygnus X-3 to
create a pair of hot spots at tens of parsec distance through
estimating its average energy injection rate. Mart\'{\i} et
al. (\cite{marti92}) have modelled the strong radio outbursts of
Cygnus X-3 in terms of laterally expanding twin jets with particle
injection at the jet basis. According to this model, a flaring event
with cm peak flux density of 10-20 Jy involves a total energy of $\sim
10^{44}$ erg in relativistic particles.  Based on several years of
radio monitoring of Cygnus X-3 with the Green Bank interferometer
(Waltman et al. \cite{w94}), two of such events occur typically per
year. Thus we find that $\sim 10^{37}$ erg s$^{-1}$ can be adopted as
a conceivable estimate. A similar value is obtained if one considers
the quiescent (low-level) radio flares of Cygnus X-3, whose energy
content is about two orders of magnitude lower and evolve within time
scales of few hours.  Remarkably, this energy injection rate is in
close accordance with the corresponding value estimated by Kaiser et
al. (\cite{kaiser04}) for the superluminal microquasar GRS 1915+105.
Such agreement strongly suggests that both microquasars could sustain
similar double lobe/hot spot features when their relativistic ejecta
interact with the ISM.

The non-thermal spectrum of the Cygnus X-3 HSCs agrees with the
expected synchrotron emission mechanism, suggesting the existence of
relativistic particles with a magnetic field. Assuming standard
equipartition arguments and a 10 kpc distance, the minimum energy
requirements and associated physical parameters can be estimated by
using the observed flux density and angular size (Longair
\cite{longair94}).

Provided that the magnetic field covers the entire radio source volume
and electrons and protons share equal energies, this implies a minimum
energy of $\sim 8 \times 10^{43}$ erg. The corresponding magnetic
field is then $3 \times 10^{-4}$ G and the total pressure $1.8 \times
10^{-9}$ erg cm$^{-3}$. These values are within a factor of a few as
compared to the non-thermal feature next to the southern lobe
candidate of GRS 1915+105 (Kaiser et al. \cite{kaiser04}).

The observed angular separation between Cygnus X-3 and its possible
hot spots transform into a linear size of 20 and 13 pc times $D_{10}
(\sin{\theta})^{-1}$ for the northern and southern candidates,
respectively.  These values are roughly a factor of three smaller than
those of GRS 1915+105. They are not expected to significatively change
even in time scales of many years. Indeed, for a strong shock in
monoatomic gas the advance velocity of the end of the jet can be
expressed as

\begin{equation}
\dot{L_j} \sim \sqrt{{16 k_B T \over 3 m_H}},
\end{equation} 

where $k_B$ is the Boltzmann constant, $m_H$ the proton mass and $T$
the temperature of the shocked gas (Kaiser et al. \cite{kaiser04}).

Assuming $T$\gtsima$10^4$ K for the post-shock temperature of the
bremsstrahlung emitting bow-shock (yet undetected for Cygnus X-3), one
expects a lobe advance velocity of \gtsima$20$ km s$^{-1}$.  This
value is well consistent with the proper motion upper limit derived in
a previous section from observations with a time baseline of several
years.  Using again the Kaiser \& Alexander (\cite{kaiser97})
formalism for a supersonically expanding lobe into a constant density
medium, the corresponding age of the jet is given by the following
self-similar solution

\begin{equation}
t = {3 \over 5} {L_j \over \dot{L_j}},
\end{equation} 

which yields an age of \ltsima$0.6$ Myr.  This upper limit
satisfactorily agrees with the expected lifetime of a Wolf-Rayet star,
which is determined by the nuclear burning time of its helium
core. For a massive binary system, the core mass is expected to be
$M_1 \simeq 8~M_{\odot}$, while the nuclear burning time of helium
depends as (Lipunov \cite{l92})

\begin{equation}
t_{\rm He} \simeq 3 \times 10^{6}{\rm yr} \left[M_1/M_{\odot} \right]^{-0.7}.
\end{equation}

This yields $t_{\rm He} \simeq 0.7$ Myr which would be satisfactorily
agree with the estimated jet age.

\section{Conclusions}

Based on a deep VLA map at the 6~cm wavelength, we have discovered two
radio sources in the vicinity of Cygnus X-3 which we propose to be the
hot spots of its relativistic jets. Assuming this statement to be
correct, the total extent of the jet flow would be measured in tens of
parsecs. The proposed physical association is based on the following
evidences:

\begin{enumerate}

\item The line joining the two HSCs agrees within one degree with the
position angle of the inner arcsecond and sub-arcsecond jets.

\item Their spectral index is consistent with a non-thermal emission
mechanism, which would naturally be expected from synchrotron
radiation in the jet bow-shocks.

\item We find no evidence for a point-like optical and near infrared
counterpart, as one would expect if the radio sources were stellar or
AGN in nature.

\item The physical parameters of the HSCs (magnetic field, pressure,
energy density, advance velocity, etc.) are comparable to those of the
proposed hot spots in GRS 1915+105. The possible existence of similar
structures in both microquasars should not come with surprise since
their estimated energy injection rate into the ISM is also very
similar ($\sim 10^{37}$ erg s$^{-1}$).

\item For Cygnus X-3, the kinematical age of the proposed hot spot
complex would be consistent with the expected lifetime of the
Wolft-Rayet stage in the binary system.

\end{enumerate}

If the hot spot nature is confirmed by further observations, the
analogy between microquasars and their extragalactic relatives would
be extended beyond the accretion/ejection point of view, and include
the interaction of their relativistic flows with the surrounding
medium, either interstellar or intergalactic.  We can even especulate
that a dichotomy, similar to that of Fanaroff-Ryle type I and II
radiogalaxies, could also exist in the microquasar domain.

\begin{acknowledgements}

The authors acknowledge support by DGI of the Ministerio de
Educaci\'on y Ciencia (Spain) under grants AYA2004-07171-C02-02 and
AYA2004-07171-C02-01, FEDER funds and Plan Andaluz de Investigaci\'on
of Junta de Andaluc\'{\i}a as research group FQM322.  The National
Radio Astronomy Observatory is a facility of the National Science
Foundation operated under cooperative agreement by Associated
Universities, Inc. in the USA. This paper is also based on
observations collected at the Centro Astron\'omico Hispano Alem\'an
(CAHA) at Calar Alto, operated jointly by the Max-Planck Institut
f\"ur Astronomie and the Instituto de Astrof\'{\i}sica de
Andaluc\'{\i}a (CSIC).  This research has made use of the SIMBAD
database, operated at CDS, Strasbourg, France. We also used the data
products from the Two Micron All Sky Survey, which is a joint project
of the University of Massachusetts and the Infrared Processing and
Analysis Center/California Institute of Technology, funded by the
National Aeronautics and Space Administration and the National Science
Foundation of USA.  The research of DPR has been supported by the
Education Council of Junta de Andaluc\'{\i}a (Spain).

\end{acknowledgements}

\end{document}